\documentclass[12pt, a4paper]{article}
\pdfoutput=1
\usepackage{amsmath,amssymb,mathrsfs}
\usepackage{graphicx, subcaption, color}
\usepackage{cancel, comment, cite, setspace}

\newcommand{\mc}[1]{\mathcal{#1}}

\definecolor{purple}{rgb}{0.5 ,0, 0.7}
\definecolor{bluegreen}{rgb}{0, 0.45, 0.35}
\usepackage{hyperref}
\hypersetup{colorlinks=true, linkcolor=bluegreen, citecolor=purple, urlcolor=blue}

\begin{document}
\begin{titlepage}
\begin{center}
\leavevmode \\
\vspace{-2 cm}

\hfill APCTP Pre2016-011\\
\hfill IPMU 16-0077

\noindent
\vskip 3 cm

\begin{spacing}{1.8}
{\LARGE \bf 
On SUSY Restoration in Single-Superfield Inflationary Models of Supergravity
}
\end{spacing}

\vglue.4in
{\large
Sergei V. Ketov~${}^{a,\, b,\, c}$ and Takahiro Terada~${}^{d}$ 
}

\vglue.3in

{\em
${}^a$~Department of Physics, Tokyo Metropolitan University \\
Minami-ohsawa 1-1, Hachioji-shi, Tokyo 192-0397, Japan \\
${}^b$~Kavli Institute for the Physics and Mathematics of the Universe (IPMU)
\\The University of Tokyo, Chiba 277-8568, Japan \\
${}^c$~Institute of Physics and Technology, Tomsk Polytechnic University\\
30 Lenin Ave., Tomsk 634050, Russian Federation \\
${}^d$~Asia Pacific Center for Theoretical Physics, Pohang 37673, South Korea
}

\vglue.1in
ketov@tmu.ac.jp, takahiro.terada@apctp.org

\end{center}

\vglue.4in

\begin{abstract}
We study the conditions of restoring supersymmetry (SUSY) after
inflation in the supergravity-based cosmological models with a single
chiral superfield and a quartic stabilization term in the K\"{a}hler
potential.
Some new, explicit, and viable inflationary models satisfying those
conditions are found.
The inflaton's scalar superpartner is dynamically stabilized during and after inflation.
We also demonstrate a possibility of having small and adjustable SUSY breaking with a tiny cosmological constant.
\end{abstract}
\end{titlepage}


\section{Introduction}

Inflation is the excellent scenario to solve the fundamental difficulties of the hot big-bang cosmology such as the horizon, flatness, and monopole problems~\cite{Starobinsky:1980te, Sato:1980yn, Guth:1980zm, Linde:1981mu, Albrecht:1982wi, Linde:1983gd}.  Moreover, it predicts generation of the curvature perturbations from quantum fluctuations of a scalar called inflaton.  Its adiabatic, scale-invariant, and Gaussian features have been precisely measured by the cosmic microwave background (CMB) observations like WMAP~\cite{Bennett:2012zja, Hinshaw:2012aka} and 
Planck~\cite{Ade:2015xua, Ade:2015lrj}. The observed small deviation from the scale invariance of CMB is
measured by the spectral index $n_{\text{s}}=0.9666\pm0.0062$~\cite{Ade:2015xua, Ade:2015lrj}, and the relative magnitude of tensor perturbations is parameterized by the tensor-to-scalar ratio $r<0.07$~\cite{Array:2015xqh}.

Inflation should be described not only phenomenologically but also consistently with particle physics expected beyond the
Standard Model. One of the most motivated approaches is supersymmetry (SUSY), or its gauged version called  supergravity~\cite{Freedman:1976xh, Deser:1976eh, Wess:1992cp, Freedman:2012zz}. In supergravity, it is known to be non-trivial to obtain a sufficiently flat inflaton scalar potential needed to trigger slow-roll inflation~\cite{Copeland:1994vg}. It is called the $\eta$ problem, and its primary cause is the presence of the exponential factor $e^{K}$ in the scalar potential (see Appendix~\ref{sec:review}).
For a small-field inflation, one may attempt to tune the parameters of the model to make the potential flat, but it becomes much more difficult for large-field inflationary models. A simple way to suppress the exponential steepness is to invoke an (approximate) shift symmetry of the K\"{a}hler potential~\cite{Kawasaki:2000yn, Kawasaki:2000ws}\footnote{
The shift symmetry may be viewed as a non-linear realization of U(1) symmetry. The U(1)-symmetric formulation of inflation in supergravity was studied in Refs.~\cite{Li:2014vpa, Li:2014unh, Li:2015mwa, Li:2015taa} with a stabilizer field, and in 
Ref.~\cite{Ketov:2015tpa} without a stabilizer field.}, though it often leads to a potential unbounded from 
below~\cite{Kawasaki:2000yn, Kawasaki:2000ws}.

Apart from tuning of the superpotential~\cite{Goncharov:1983mw, Roest:2015qya, Linde:2015uga},
there are two known generic solutions to the unboundedness problem.
The first one is to introduce a stabilizer superfield $S$ whose value is required to vanish on-shell~\cite{Kawasaki:2000yn, Kawasaki:2000ws, Kallosh:2010ug, Kallosh:2010xz}.
The superpotential is taken to be proportional to that superfield, so that the negative contribution $-3|W|^2$ is effectively removed.  This approach can also be used by assuming the $S$ to be a nilpotent superfield, thus effectively eliminating the need of its stabilization and invoking non-linear realizations of local supersymmetry and its breaking~\cite{Antoniadis:2014oya, Ferrara:2014kva, Kallosh:2014via, Dall'Agata:2014oka, Dall'Agata:2015lek}, see \textit{e.g.}  Refs.~\cite{Dall'Agata:2016yof, Kallosh:2016hcm, Ferrara:2016een} and references therein for more recent 
contributions.  
The other approach is to introduce a (shift-symmetric)\footnote{
A model without shift symmetry was proposed in Ref.~\cite{Ketov:2014qha}. It is related to the single-superfield $\alpha$-attractor proposed in Refs.~\cite{Roest:2015qya, Linde:2015uga, Scalisi:2015qga}. See also Refs.~\cite{Carrasco:2015uma, Terada:2015sna} for more.
} quartic term of the inflaton in the K\"{a}hler potential~\cite{Izawa:2007qa, Ketov:2014qha, Ketov:2014hya, Ketov:2015tpa}, instead.\footnote{
Some tuned and complicated K\"{a}hler potentials with a similar stabilization mechanism were used in Ref.~\cite{Izawa:2007qa}. 
Their (shift-symmetric) simplifications were proposed in Refs.~\cite{Ketov:2014qha, Ketov:2014hya} where the quartic  stabilization mechanism was pointed out to be applicable to generic inflaton potentials with appropriate superpotentials. As regards further
developments of this approach, see Refs.~\cite{Linde:2014ela, Gao:2015yha, Terada:2015cna, Nastase:2015pua}.
The importance of K\"{a}hler curvature (represented by a quartic term in our K\"{a}hler potential) was emphasized in Ref.~\cite{Covi:2008cn}.
} 
 The quartic term plays the role of a SUSY breaking mass term for the scalar superpartner of the inflaton (we call it
{\it sinflaton}). It also has another important (dual) role by lifting up the inflaton potential to make it positive, and fixing the value of the sinflaton during inflation, in order to make inflationary dynamics to be the single-field one (not just the 
single-\emph{super}field one). 
 The latter approach reduces the matter degrees of freedom (needed for inflation and supersymmetry) to half of the former approach with a stabilizer (super)field.  In this sense, it is a minimal approach to (large-field) inflation in supergravity.
Therefore, it is worth studying the properties of those relatively new supergravity-based inflationary models in more detail.

Generic models with a single chiral superfield (in other words, without a stabilizer) were introduced in Ref.~\cite{Ketov:2014qha}, where it was found that SUSY is generically {\it not} restored after inflation. Hence, without a hierarchically small parameter, SUSY tends to be broken in vacuum at a scale comparable to the inflation scale, \textit{i.e.} the gravitino mass is approximately of the same order as the inflaton mass. The conditions to restore SUSY after inflation were not addressed in Ref.~\cite{Ketov:2014qha}. Besides, in Ref.~\cite{Ketov:2014hya}, we introduced the special logarithmic 
K\"{a}hler potential that allows an approximate embedding of \emph{arbitrary} positive semi-definite scalar potentials.  In that class of models, it is also possible to fine-tune the cosmological constant and SUSY breaking after inflation to zero. So,
it can be a good starting point of the model building to obtain a small positive cosmological constant and SUSY breaking that would be parametrically smaller than the inflation scale. It is also worth investigating whether this feature is maintained after taking into account corrections of the order $1/\zeta$, with $\zeta$ being the strength of the quartic stabilisation term. 
 Whether SUSY is restored after inflation (in the absence of the hidden SUSY breaking sector) is quite important, being
related to the grand unification of the gauge coupling constants and to the gauge hierarchy problem of the Standard Model of particle physics. Should SUSY be broken at a scale higher than the intermediate scale, the electroweak vacuum may be unstable~\cite{Degrassi:2012ry}. Moreover, gravitino production from inflaton decay in the early Universe is enhanced when inflaton breaks SUSY in vacuum~\cite{Endo:2006zj, Nakamura:2006uc, Kawasaki:2006gs, *Kawasaki:2006hm, Asaka:2006bv}, which leads to a cosmological disaster.

In this paper, we study SUSY breaking (and its preservation) in vacuum, by using the supergravity setup utilising a quartic stabilisation term in the K\"{a}hler potential. More specifically, we study the conditions to restore SUSY after inflation.
We find that SUSY restoration is intact in the presence of finite corrections in $1/\zeta$. In Section~\ref{sec:general}, we discuss a general setup. A specific model is introduced and studied in Section~\ref{sec:analyses}. We find a new 
two-parametric generalization of the $\alpha$-attractor potential~\cite{Ellis:2013nxa, Ferrara:2013rsa, Kallosh:2013yoa}, which leads to slightly different predictions compared to the original $\alpha$-attractor's. 
In Section~\ref{sec:breaking} we outline how to get an adjustable cosmological constant with SUSY breaking.
Section~\ref{sec:conclusion} is our Conclusion.
 Our setup is described in Appendix~\ref{sec:review}.  Various models which restore SUSY after inflation are presented in Appendix~\ref{sec:models}. 
We adopt the natural units, $c=\hbar=M_{\text{Pl}}/\sqrt{8\pi}=1$.

\section{SUSY restoration in a generic case}\label{sec:general}
We consider the ``generic''\footnote{
For instance, the K\"{a}hler potential of Ref.~\cite{Ketov:2014hya} --- see Eq.~\eqref{Ksp} --- can be approximately expressed by Eq.~\eqref{K} as a Taylor series.  Then the linear coefficient is $c=-\sqrt{3}$, the quartic coefficients are related as $\zeta = \xi - 1/3$, and a small cubic term, $i(\Phi-\bar{\Phi})^3/(3\sqrt{3})$ is needed, whose presence merely results in subdominant effects.
}  shift-symmetric K\"{a}hler potential of Ref.~\cite{Ketov:2014qha}.
 We take the convention that the real part $\phi$ of the leading component of a chiral superfield, 
$\left.\Phi\right|=(\phi+ i \chi)/\sqrt{2}$, defines the shift-symmetric direction, \textit{i.e.}
\begin{align}
K= i c \left( \Phi - \bar{\Phi}\right) -\frac{1}{2} \left( \Phi - \bar{\Phi}\right) ^2 -\frac{\zeta}{4} \left( \Phi - \bar{\Phi}\right)^4, \label{K}
\end{align}
where $c$ is a real constant and $\zeta$ is a real positive constant. Inflaton is identified with $\phi$ that enters a superpotential.
The sinflaton $\chi$ is stabilized 
 by the quartic term during inflation~\cite{Ketov:2014qha, Terada:2015sna}.
We choose the origin of sinflaton $\chi$ in such a way that it coincides with its stabilized value, \textit{i.e.} $\langle \chi \rangle \to 0$ as $\zeta \to \infty$. 
That is why the cubic term in Eq.~\eqref{K}, which would induce unsuppressed $\langle \chi \rangle$, is assumed to be negligible.
With this choice, the higher order terms in $i(\Phi - \bar{\Phi})$ have negligible effects because of the suppression $\langle \chi \rangle \simeq 0$.
It is worth mentioning that these choices are different from those in Ref.~\cite{Ketov:2014qha}, though being equivalent via a field redefinition.
The quartic term is needed only for the stabilization of the sinflaton and, hence, the related uplifting of the inflaton potential via the linear term.
During inflation, the sinflaton $\chi$ can be integrated out, 
 while the impact of the quartic term results in the appearance of the terms
 {\em inversely} 
proportional to $\zeta$ in the effective single-field potential for
inflaton $\phi$, due to the sinflaton value $\langle \chi \rangle \sim
\zeta^{-1}$.
This is because the quartic term itself is multiplied by the power of
the sinflaton value, and vanishes in the limit $\zeta\to \infty$.
Though we take into account the corrections in $1/\zeta$ in our calculations of the inflationary observables, essential features of our models can be already seen in the leading (zeroth) order in $1/\zeta$.

The effective single-field inflaton potential (in the leading order) is given by
\begin{align}
V= \left| W_{\Phi} \right|^2 + i c \left( W \bar{W}_{\bar{\Phi}} -\bar{W}W_{\Phi} \right) + (c^2-3)|W|^2.
\end{align}
Note that the coefficient of the last term is only positive for $c>\sqrt{3}$.
In the presence of corrections of the order $1/\zeta$, the critical value of the lower bound increases, as is shown below.
When assuming for simplicity that all parameters in the superpotential are real, the second term in the $V$ given above vanishes, so we get a simplified formula,
\begin{align}
V= \left| W_{\Phi} \right|^2 + (c^2-3)|W|^2. \label{Vleading}
\end{align}

The requirement of SUSY preservation in vacuum with the vanishing cosmological constant is not a severe condition, since it merely requires
\begin{align}
W=W_{\Phi}=0  &  & \text{(in vacuum).}\label{SUSYP}
\end{align}
This leads to the vanishing $F$-term, $D_{\Phi}W=W_{\Phi}+K_{\Phi}W=0$, and $V=0$.
(We discuss small SUSY breaking and a cosmological constant in Section~\ref{sec:breaking}.)
As follows from Eqs.~\eqref{K} and \eqref{SUSYP}, it is self-consistent to assume $\chi=0$ \emph{exactly} in vacuum \emph{for arbitrary $c$ and $\zeta$}.
Namely, 
\begin{align}
V=&V_{\phi}=V_{\chi}=0  &   &(\text{in vacuum with } \chi=0), \label{stationarity}
\end{align}
with the SUSY mass squared for $\phi$ and $\chi$ given by $|W_{\Phi \Phi}|^2$ (\textit{i.e.} no tachyon). 
Both $\phi$ and $\chi$ are canonically normalized at $\chi=0$, including the vacuum state.
Those facts simplify our analysis.
Moreover, they prevent large field excursion of $\chi$ at the final
phase of inflation, as observed in Ref.~\cite{Linde:2014ela}, because
the relevant expectation values (both in vacuum and during inflation)
are close to each other.

In summary, we established the following statement: once one constructs an inflationary model in the ideal stabilization limit $\zeta\to \infty$ in such a way that the vacuum is at $\phi=\phi_0$ (here $\phi_0$ is a constant) and $\chi=0$, the vacuum position is {\it unchanged} when $\zeta$ becomes finite.  
In fact, this is valid for general shift-symmetric K\"{a}hler potentials, $K(\Phi, \bar{\Phi})=K(i (\Phi-\bar{\Phi}))$.
It is straightforward to show that Eq.~\eqref{stationarity} holds under the assumption~\eqref{SUSYP}.
In the general case, the squared mass of $\phi$ and $\chi$ is given by $e^{K(0)}|W_{\Phi \Phi}|^2/K''(0)$, where the primes denote the differentiation with respect to the given argument. Note that the value of $\chi=0$ in a SUSY vacuum is not guaranteed by the quartic stabilization term because the SUSY breaking mass stabilizing $\chi$ vanishes in a SUSY vacuum.

The important corollary of our statement exists for a class of the inflationary models with the special K\"{a}hler potential
\begin{align}
K=-3 \log \left(1+ i (\Phi - \bar{\Phi})/\sqrt{3} + \xi (\Phi - \bar{\Phi})^4/12 \right)~. \label{Ksp}
\end{align}
It makes possible to incorporate an arbitrary positive semi-definite inflaton potential via the formula
\begin{align}
V=& |W_{\Phi}|^2 + \sqrt{3}i \left( \bar{W}W_{\Phi}+W\bar{W}_{\bar{\Phi}}\right) \nonumber \\
=& |W_{\Phi}|^2,
\end{align}
where the second equality holds when one assumes all the coefficients in the superpotential to be real.
It was shown in Ref.~\cite{Ketov:2014hya} that one can always fine-tune both a cosmological constant and SUSY breaking 
in vacuum to zero in the infinite $\xi$ limit. The above statement generalizes it to the case of arbitrary $\xi$.

\section{An example}\label{sec:analyses}
Let us study in more detail the conditions~\eqref{SUSYP} for SUSY preservation and the vanishing cosmological constant.
By shifting $\Phi$ with a real constant, we can set $\phi_0=0$ without changing the form of the K\"{a}hler potential.
Without loss of generality, we set it first and study the structure of the superpotential satisfying Eq.~\eqref{SUSYP}.
Then the vacuum is at the origin $\Phi=0$.  
The superpotential can be written down as a Taylor expansion without the zeroth and first order terms,
\begin{align}
W= \sum_{n \geq 2} c_n \Phi^n. \label{Wexpansion}
\end{align}
This satisfies Eq.~\eqref{SUSYP}, while any holomorphic function satisfying Eq.~\eqref{SUSYP} can be expressed as above. Given an even function $W(-\Phi)=W(\Phi)$ satisfying $W_{\Phi}=0$ at the origin, it is always possible to subtract a constant to obey the remaining condition $W=0$.

We focus on a particular example in this Section, though there exist many models with SUSY restoration after inflation.
Their possible classification within our approach is outlined in Appendix \ref{sec:models}.

\begin{figure}[htb]
 \centering
 \subcaptionbox{The examples of the potential~\eqref{Vexps}.\label{sfig:exps}}
{\includegraphics[width=0.47\columnwidth]{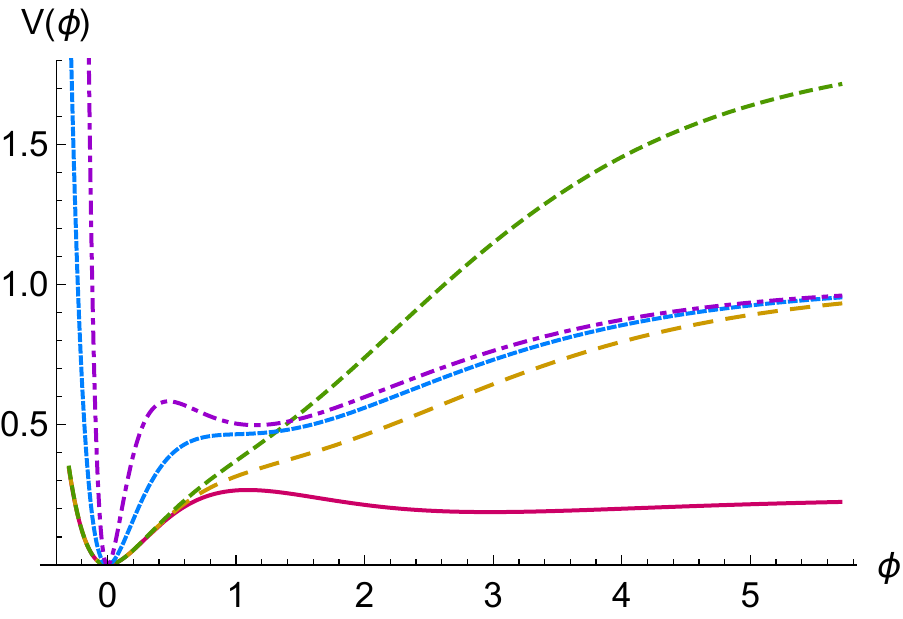}}~
 \subcaptionbox{The limit $b_2 \to b_1$ in Eq.~\eqref{defStaro}. \label{sfig:defStaro}}
{\includegraphics[width=0.47\columnwidth]{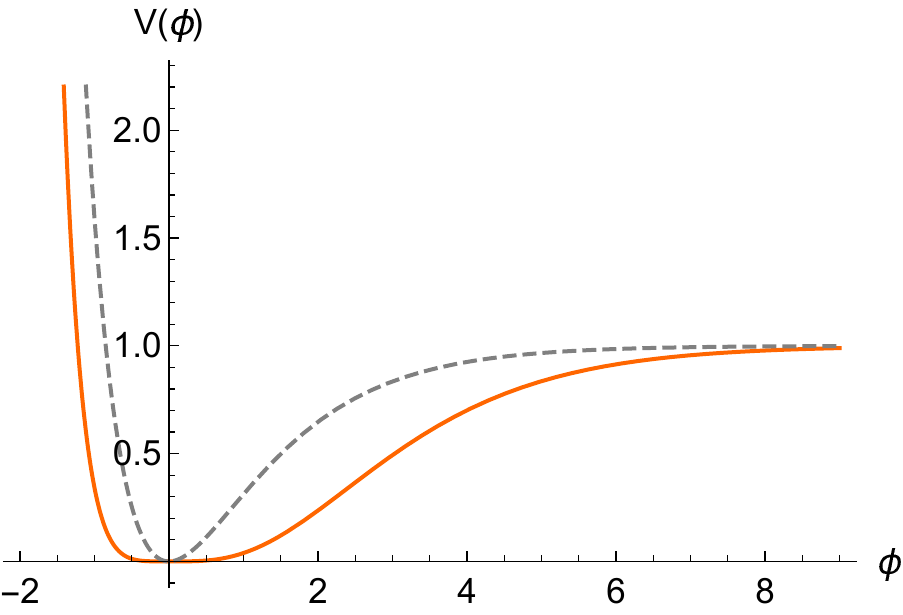}}~

  \caption{The examples of the potential in the model~\eqref{racetrack}. 
 (\ref{sfig:exps}) the parameters are chosen to be $(a_0, b_1, b_2, c)= (1,1,2,1.8), (1,1,2,2), (1, 1, 2, 2.2), (1, 1, 4, 2),$ and 
 $(1, 1, 8, 2)$ for the red solid, yellow long-dashed, green dashed, blue short-dashed, and purple dot-dashed lines, respectively. (\ref{sfig:defStaro}) the potential~\eqref{defStaro} with $b=\sqrt{2/3}$.  The Starobinsky potential is shown by the dashed line for comparison.
  }
 \label{fig:exps}
  \end{figure}

The CMB data favours a flat potential similar to the potential of the $R^2$ model~\cite{Starobinsky:1980te} or Higgs inflation model~\cite{Bezrukov:2007ep}. It is, therefore, reasonable to consider a potential that asymptotes to a constant at the
infinite inflaton field, $\phi\to\infty$. An asymptotically constant potential is generated by an asymptotically constant superpotential. Such a superpotential can be expanded as 
\begin{align}
W = \sum_{n\geq 0}  a_n  e^{-b_n \Phi}~~,
\end{align}
where $b_0=0$ and $b_n> b_m$ for $n>m$.
The condition \eqref{SUSYP} then reads
\begin{align}
\sum_{n\geq 0} a_n =& 0  & \text{and}& &
\sum_{n\geq 1} a_n b_n =& 0~.  \label{SUSYPexp}
\end{align}
It does not have a non-trivial solution when there are only two terms in the expansion. Hence,
let us consider the simplest non-trivial case with
\begin{align}
W= a_0 + a_1 e^{-b_1 \Phi} + a_2 e^{-b_2 \Phi}. \label{racetrack}
\end{align}
This superpotential is the same as that in the so-called racetrack model~\cite{Escoda:2003fa}, though our K\"{a}hler potential is different. 
By solving the constraint~\eqref{SUSYPexp}, we can eliminate $a_1$ and $a_2$ as
\begin{equation}
a_1 = -\frac{a_0 b_2}{b_2 - b_1}    \quad  {\rm and} \quad a_2= \frac{a_0 b_1}{b_2 - b_1}~~.
\end{equation}
Then the potential~\eqref{Vleading} in the leading order takes the form
\begin{align}
V=& \frac{a_0^2}{(b_2 - b_1)^2}\left[ b_1^2 b_2^2\left(e^{-\tilde{b}_1 \phi}-e^{-\tilde{b}_2 \phi} \right)^2 + (c^2-3) \left( b_2 - b_1 + b_1 e^{-\tilde{b}_2 \phi} - b_2 e^{-\tilde{b}_1 \phi} \right)^2 \right], \label{Vexps}
\end{align}
where $\tilde{b}_i=b_i /\sqrt{2}$ $(i=1, 2)$.
Some visual examples of the potential are shown in Fig.~\ref{fig:exps}.

A shape of the contribution to the potential, coming from the derivative term $W_{\Phi}$, is not suitable for a plateau inflation. In fact, when $c$ is closer to the critical value $\sqrt{3}$ and/or $b_1$ and $b_2$ are large (the rough criterium is $b_1^2b_2^2 \gtrsim (c^2-3)(b_2-b_1)^2$), 
a bump in the potential appears.  The red solid and purple dot-dashed lines in  Fig.~\ref{sfig:exps} are such examples.

To study the $b_i$-dependence in our models, we consider the large $c$ limit in which the first term in Eq.~\eqref{Vexps}, originating from the derivative of the superpotential, is negligible. When $b_2$ becomes much larger than $b_1$, the terms with the higher-oder exponentials also become negligible.
 It means that the potential is Starobinsky-like. More precisely, it coincides with the E-model-type realization 
of the $\alpha$-attractor potential~\cite{Kallosh:2015lwa}. Actually, a large $b$ with a fixed $c$ implies that the neglected term becomes important. When $b_2$ is closer to $b_1$, the situation is more complicated. The most nontrivial case is the limit $b_2 \to b_1$. Then, the superpotential and the potential become
\begin{align}
W= & a_0 \left[ 1 -(1+ b_1 \Phi) e^{-b_1 \Phi} \right] ,\\
V=& a_0^2 (c^2 -3) \left[ 1 - (1 + b \phi) e^{-b \phi} \right]^2, \label{defStaro}
\end{align}
where $b\equiv b_1 / \sqrt{2}$.
This potential is different from the Starobinsky one, as is shown in Fig.~\ref{sfig:defStaro}.
In this limit, the neglected derivative-originated term is $\sqrt{2} a_0^2 b^2 \phi e^{-b \phi}$, so that the above approximation is valid for $b\lesssim c$.
As a rough estimate of the inflationary observables for these cases including the Starobinsky-like limit, we have 
\begin{align}
n_{\text{s}}\simeq 1- \frac{1}{2N},  \quad  {\rm and} \quad r \simeq \frac{8}{b^2 N^2}.
\end{align}

In our numerical calculations we took into account a shift of the sinflaton $\chi$ from the origin up to the first order.
For this purpose, we expanded the potential in terms of $\chi$ up to the second order and minimized it.
After integrating out $\chi$, we obtained the effective single-field potential of the inflaton $\phi$, including the corrections 
having the $\zeta$-dependent terms. So, we neglected the time derivatives of $\chi$. To do those calculations efficiently, we approximated the potential by an interpolation method of Mathematica, and solved the equation of motion to obtain the inflaton trajectory as well as the inflationary observables. For a reduction of the multi-parameter space, we set the stabilization parameter to be $\zeta=1$. The overall scale of the potential merely affects the time-scale of simulations, so we set $a_0=1$.

\begin{figure}[htb]
 \centering
\includegraphics[width=0.8\columnwidth]{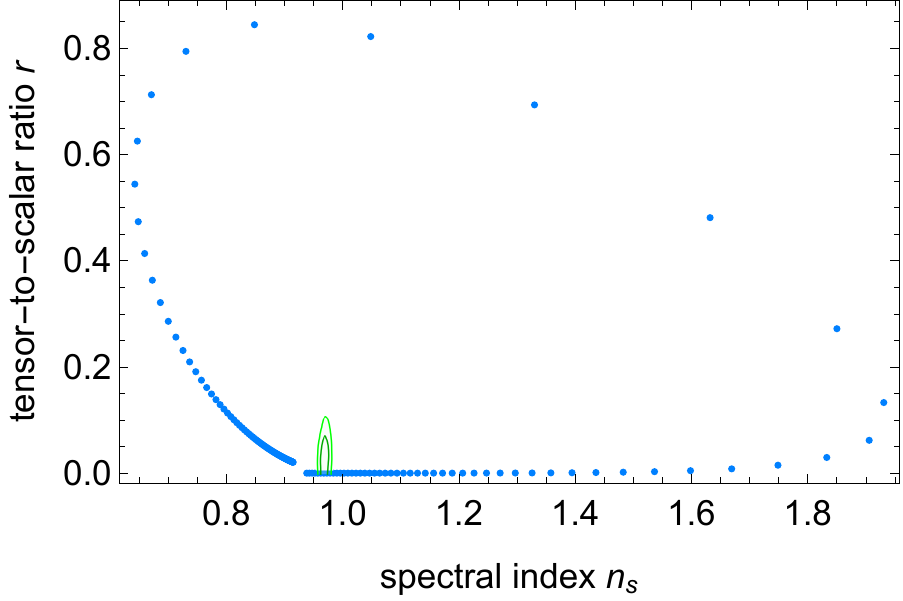}
  \caption{A rapid change of predictions near the critical value of the parameter $c$.
  The value of $c$ is varied from 1.98359 to 1.98500 with a 0.00001 step each.
  The points circulate counter-clockwise with increasing $c$.
  The other parameters are set to $\zeta=a_0=1$, $b_1=2/\sqrt{3}$, $b_2=2 b_1$, and $N=60$.
  }
 \label{fig:c-dep}
  \end{figure}

With those corrections, the critical value of $c$, needed to obtain a potential bounded from below, increases a bit. The critical value of $c$ is also dependent on the other parameters, in particular $b_1$ and $b_2$.
One may also study the critical values of $b_i$ at a fixed $c$. Near their critical values, the effect of the derivative-induced term is not negligible. When decreasing $c$ from a large value to the critical value, the potential around the origin is gradually deformed, a short flat region appears, and finally it becomes a bump to trap inflaton into a local minimum.
Some of these features can be seen in Fig.~\ref{sfig:exps}. Around that small parameter range of $c$, the $e$-folding number earned in the short flat region changes rapidly, and it is reflected in a rapid change of the corresponding predictions for $(n_{\text{s}}, r)$. This is demonstrated in Fig.~\ref{fig:c-dep} where one looses predictability against flexibility of  predictions in this very special case.
 
 \begin{figure}[htb]
 \centering
\includegraphics[width=0.8 \columnwidth]{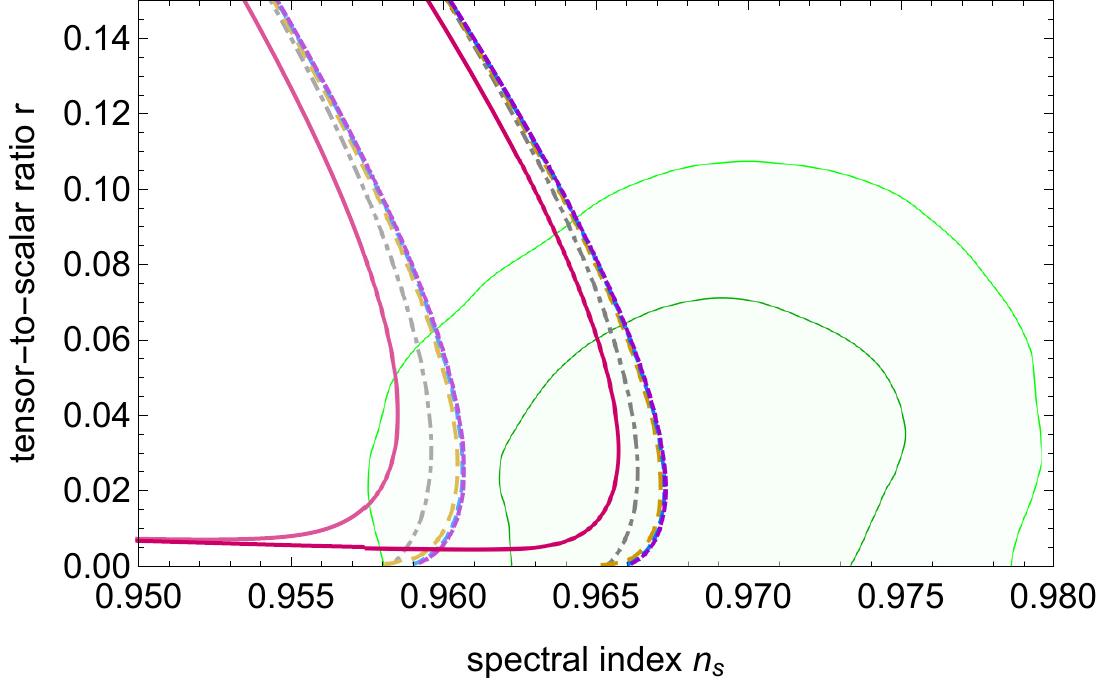}
  \caption{The $b_i$-dependence in the model with two exponentials~\eqref{racetrack}.
  The set of right (left) lines with darker (lighter) colors corresponds to $N=60$ ($50$).
  In each set, the red solid (the very left), yellow long-dashed, blue dashed, and purple short-dashed (the very right) lines show the cases of $c=2, 3, 4$, and $5$, respectively.  The lines with $c=3, 4$, and $5$ are almost degenerate.
  We set $b_2 = 2 b_1$ and $\zeta=a_0=1$, and vary $b_1$ from 0.01 above in the figure to 3.0 (1.2 for $c=2$) below in the figure. For comparison, the result of the $b_2 \to b_1$ limit with large $\zeta$ and $c$ (see 
  Eq.~\eqref{defStaro}) is shown as the gray dot-dashed line.
  The green contours are $1\sigma$ and $2\sigma$ Planck constraints combined with other observations (Planck~TT+lowP+BKP+lensing+BAO+JLA+H0)~\cite{Ade:2015xua}.
  }
 \label{fig:b-dep}
  \end{figure}
 
For larger values of $c$, observational predictions vary less rapidly, as usual. The predictions of our models are shown in 
Fig.~\ref{fig:b-dep} where we have fixed the relation $b_2 = 2 b_1$ for definiteness, and have taken $c=2, 3, 4$, and $5$.
The value of $b_1$ is varied from 0.01 to 3.0 at $c=3, 4$, and $5$. When $c=2$, the maximum value is taken to be 1.2
that is almost the critical value for $c=2$. For this reason, only the red solid line for $c=2$ is deflected to the lower $n_{\text{s}}$ region.

\section{SUSY breaking and cosmological constant} \label{sec:breaking}
It is worthwhile to comment on the related issue, as to how to obtain {\it small} SUSY breaking and a {\it very small} cosmological constant, by some minimalistic extension of our approach. For simplicity, let us consider a nilpotent $F$-term SUSY breaking superfield $S$~\cite{Ferrara:2014kva}, which is essentially a Polonyi superfield~\cite{Polonyi:1977pj} subject to the nilpotent condition $S^2=0$.\footnote{
One may impose further constraints, $\bar{S}S (\Phi - \bar{\Phi}) = 0$ and $\bar{S}S \mc{D}_{\alpha} \Phi = 0$, to eliminate sinflaton and inflatino, respectively~\cite{Dall'Agata:2016yof}.
} We take the minimal K\"{a}hler potential of the nilpotent superfield, and combine the two sectors as follows:
\begin{align}
K(\Phi, S, \bar{\Phi},\bar{S})=& K^{(\text{inf})}(i(\Phi - \bar{\Phi})) +\bar{S}S, \\
W(\Phi, S)=& W^{(\text{inf})}(\Phi) + W_0 + \mu^2 S ,
\end{align}
where $W_0$ and $\mu^2$ are constants, and (inf) denotes the quantities in the inflation sector discussed in this paper.
In particular, $W^{(\text{inf})}$ satisfies Eq.~\eqref{SUSYP}, and we set $K^{(\text{inf})}(0)=0$ as our convention by using 
a K\"{a}hler transformation. In the presence of the superfield $S$, the VEV of the inflaton gets shifted as
\begin{align}
\langle \Phi \rangle = -  \frac{W_0 K^{(\text{inf})}_{\Phi}}{W^{(\text{inf})}_{\Phi\Phi}}~~,
\end{align}
where we have taken the convention $\langle \Phi \rangle = 0$ before introducing $W_0$ and $S$, and the terms of the higher order in $|W_0|$ or $|\mu|^2$ have been neglected. Accordingly, the vacuum energy becomes
\begin{align}
V= |\mu|^4 - 3 |W_0|^2,
\end{align}
in the leading order of $|W_0|$ or $|\mu|^2$.
Therefore, the SUSY breaking scale $|D_S W|=|\mu^2|$ can be chosen freely, while the cosmological constant can be 
chosen arbitrarily small by fine-tuning between $|\mu|^4$ and $3|W_0|^2$. For many purposes, we may simply set 
$|\mu|^2\approx \sqrt{3}|W_0|$, so that the gravitino mass is  $m_{3/2}=|W_0|=|\mu^2|/\sqrt{3}$.

The role of the superfield $S$ in our approach is limited to uplifting
the vacuum energy.
The inflaton potential is solely constructed from the inflaton superfield $\Phi$.
If the SUSY breaking scale is much lower than the inflation scale, the effects of $S$ or $W_0$ on the inflationary dynamics are negligible.

Our results are consistent with the argument in Ref.~\cite{Linde:2014ela} and the ``no-go'' statement in Ref.~\cite{Kallosh:2014oja}, which claim that any SUSY preserving Minkowski vacuum without flat directions {\it cannot} be uplifted to a de Sitter vacuum by a small continuous deformation of the model. However, there exists a loophole in those arguments, also noticed in Ref.~\cite{Kallosh:2014oja}, namely, via adding a massless superfield (a flat direction) and increasing its mass, i.e. exactly as we did above.

\section{Conclusion}\label{sec:conclusion}

In this paper we investigated the SUSY breaking properties of the supergravity-based inflationary models without a stabilizer superfield by using a shift-symmetric quartic stabilization term in the K\"{a}hler potential. 
The shift symmetry is a global symmetry imposed on the K\"{a}hler potential at the tree level, which is likely to be broken by quantum (gravity) corrections. Though quantum corrections from the superpotential are suppressed due to the relatively small scale controlling the amplitude of CMB perturbations, one may expect non-negligible quantum corrections from the inflaton-matter couplings, depending upon the reheating temperature. We assumed those terms to be suppressed in our phenomenological approach. Possible origins of our superpotentials and K\"{a}hler potential, and, in particular, measuring quality of the shift symmetry, are beyond the scope of our investigation.

In Section~\ref{sec:general}, we found that the vacuum expectation value (VEV) of the inflaton multiplet is not sensitive to the parameters of the K\"{a}hler potential as far as the condition~\eqref{SUSYP} is satisfied. This demonstrates robustness of the SUSY preservation property in our models.
In addition, we showed that the large-field excursion of sinflaton at the end of inflation, observed in 
Ref.~\cite{Linde:2014ela}, can be suppressed by tuning the sinflaton VEV to be equal to its stabilized value during inflation (\textit{i.e.} zero in our conventions).

A relatively simple, racetrack-like model was studied in Section~\ref{sec:analyses}. It shares the essential qualitative features with some other models in Appendix~\ref{sec:models}. The observational aspects of the model extend those of the $\alpha$-attractor, including the $R^2$ model and the Higgs inflation --- see
\textit{e.g.}, Eqs.~\eqref{Vexps} and \eqref{defStaro}, and Figs.~\ref{sfig:defStaro} and \ref{fig:b-dep} for details.

In summary, our single-superfield model building with the quartic stabilization is a powerful tool to construct inflationary models in supergravity, which are consistent with observations. Its inflationary sector has the minimal number of physical degrees of freedom, \textit{i.e.} has the inflaton supermultiplet only. 
Since sinflaton is stabilized, isocurvature perturbations and non-Gaussianity are negligible in our models, see \textit{e.g.}, Ref.~\cite{Hetz:2016ics} for more.
The SUSY breaking by the inflaton supermultiplet driving inflation is restored after inflation when the condition~\eqref{SUSYP} is satisfied.
On top of that, we found that it is possible to obtain a tunable SUSY breaking and a tiny cosmological constant in vacuum.
It fills a gap in our earlier work on the single-superfield approach to inflation in supergravity, as regards its SUSY breaking structure after inflation.

\section*{Acknowledgements}
The authors are grateful to H.~Murayama for raising the issue of the $1/\zeta$-corrections to the SUSY breaking properties, 
which has led us to our investigation in Section~\ref{sec:general}.
SVK is supported by a Grant-in-Aid of the Japanese Society for Promotion of Science (JSPS) under No.~26400252, a grant of the  President of Tokyo Metropolitan University, the World Premier International Research Center Initiative (WPI Initiative), MEXT, in Japan, and the Competitiveness Enhancement Program of the Tomsk Polytechnic University in Russia.
A part of TT's work was supported by the Grant-in-Aid for JSPS Fellows and the Grant-in-Aid for Scientific Research on Scientific Research No.~26$\cdot$10619.

\appendix

\section{Basic facts about inflationary model building} \label{sec:review}
The most important inflationary observables are (i) the amplitude of the curvature perturbations $A_{\text{s}}$\,, (ii) the scalar 
spectral index $n_{\text{s}}$\,, and (iii) the tensor-to-scalar ratio $r$.  They can be expressed in terms of the slow-roll parameters as follows:
\begin{align}
A_{\text{s}}=&\frac{V}{24\pi^2 \epsilon}  ~,  &   n_{\text{s}}=& 1- 6\epsilon + 2\eta~,   &    r=& 16 \epsilon~,
\end{align}
at the horizon exit, in terms of the inflaton scalar potential $V$. The slow-roll parameters are defined as
\begin{align}
\epsilon =& \frac{1}{2}\left(\frac{V'}{V}\right)^2 ~, &   \eta = & \frac{V''}{V}~,
\end{align}
where the primes denotes the differentiation with respect to the canonical inflaton field $\phi$.
The $e$-foldings number $N\equiv \log (a_{\text{end}}/a_{*})$ can be expressed in terms of the inflaton field as
\begin{align}
N=\int _{\phi_{\text{end}}}^{\phi_{*}} \frac{1}{\sqrt{2\epsilon}} \mathrm{d}\phi~,
\end{align}
where $a$ is the scale factor of the FLRW metric, the subscript ``end'' denotes the end of inflation 
(at $\epsilon=1$),  the subscript $(*)$ stands for the horizon exit of the observed scale, and we set $\phi_{\text{end}}<\phi_{*}$ without loss of generality.

In four-dimensional $\mc{N}=1$ supergravity, an inflationary model is specified by  a K\"{a}hler potential $K=K(\phi^i, \bar{\phi}^{\bar{j}})$, a superpotential $W=W(\phi^i)$, and a gauge kinetic function $h_{AB}=h_{AB}(\phi^i)$ of chiral
superfields $\phi^i$.  The kinetic and potential terms of their leading scalar field components 
$\phi^i$ and $\bar{\phi}^{\bar{j}}$ (where a bar denotes complex conjugation, and we use the same notation for chiral
superfields and their leading field components) in Einstein frame are given by
\begin{align}
\sqrt{-g}^{-1}\mc{L}_{\text{kinetic}}=& - K_{i\bar{j}}\partial^{\mu}\phi^i \partial_{\mu}\bar{\phi}^{\bar{j}}~, \\
V=& e^K \left( K^{i\bar{j}}D_i W \bar{D}_{\bar{j}}\bar{W} -3 |W|^2 \right) + \frac{1}{2}h_{AB}^{\text{R}} D^{A}D^{B}~,
\end{align}
where the subscripts $i$, $\bar{j}$, etc. denote the differentiation with respect to the corresponding fields 
$\phi^i$, $\bar{\phi}^{\bar{j}}$, etc., and $D_iW \equiv W_i + K_i W$.
The $D$-term (proportional to $D^A D^B$) is irrelevant for our investigation in this paper.

The minimal K\"{a}hler potential $K=\bar{\phi}\phi$ leads to a scalar potential having the overall exponentially steep factor 
$e^{\bar{\phi}\phi}$ in large-field inflationary models. A detailed review of the $\eta$-problem in supergravity can be found,
 \textit{e.g.}, in Ref.~\cite{Yamaguchi:2011kg}.

\section{Towards a classification of inflationary models in supergravity with SUSY restoration} \label{sec:models}
\paragraph{Type 1a: the single-term-model} \mbox{}\\
The simplest option is merely a single term in Eq.~\eqref{Wexpansion},
\begin{align}
W=c_2 \Phi^2.
\end{align}
Then the leading-order scalar potential is a sum of quadratic and quartic terms,
\begin{align}
V=\frac{|c_2|^2}{4} \phi^2 \left[ 8 + \left( c^2 -3\right) \phi^2 \right].
\end{align}
However, both quadratic and quartic potentials are too steep to be consistent with observations.
Tensions with observations become milder when a small {\it negative} quartic term is added to a quadratic potential~\cite{Buchmuller:2015oma}. Though the potential then becomes unbounded from below in the large-field limit, it may not 
be a problem if the tunnelling time  is longer than the age of the Universe.

\begin{figure}[htb]
 \centering
\includegraphics[width=0.6\columnwidth]{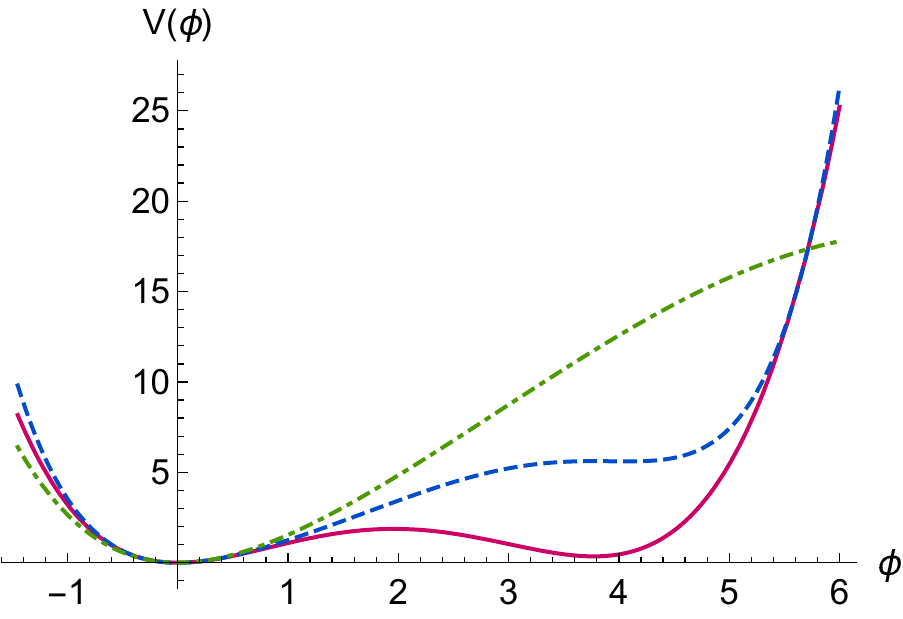}
  \caption{The examples of the potential~\eqref{pot_2terms}.
  The parameters are chosen as $(c, c_3)= (1.75, -0.25), (2.0, -0.25),$ and $(1.8, -0.13)$ for the red solid, blue dashed, and green dot-dashed lines, respectively. In all cases we set $c_2 =1$.
  }
 \label{fig:2terms}
  \end{figure}

\paragraph{Type 1b: the two-terms-model} \mbox{}\\
After adding a cubic term to the previous model, we get
\begin{align}
W=c_2 \Phi^2 + c_3 \Phi^3.
\end{align}
The potential in the leading order is polynomial,
\begin{align}
V= 2c_2^2 \phi^2+3\sqrt{2}c_2 c_3 \phi^3 + \frac{9}{4}c_3^2 \phi^4 + \frac{1}{8}(c^2-3)\phi^4 \left( 2 c_2^2+2\sqrt{2}c_2 c_3 \phi + c_3^2 \phi^4 \right), \label{pot_2terms}
\end{align}
where we have taken $c_2$ and $c_3$ to be real for simplicity.
Some examples of such potential are shown in Fig.~\ref{fig:2terms}.
In the limit of $c=\sqrt{3}$, the potential is a quartic function and has the double-well form.
The positions of the minima (with the vanishing cosmological constant) are $\phi_0=0$ and $\phi_0=-2\sqrt{2}c_2 / (3 c_3)$.
SUSY is preserved in the former minimum and is broken in the latter. A hilltop inflation of the quartic order is possible between the minima, but it gives $n_{\text{s}}\simeq 0.94 \sim 0.95$ which is smaller than the observational bound.
When we increase $c$, the nontrivial minimum is uplifted, but the local minimum still exists. 

In the two examples above, we merely considered the simplest options. Next, we require that the potential asymptotes 
to a constant in the large-field limit (with some values of the parameters). The leading-order potential~\eqref{Vleading} 
has two parts: the derivative part and the non-derivative part. Accordingly, there exist two possibilities where one of the two parts becomes dominant.

\begin{figure}[htb]
 \centering
 \subcaptionbox{The asymptotically linear superpotentials.\label{sfig:aLinearW}}
{\includegraphics[width=0.47\columnwidth]{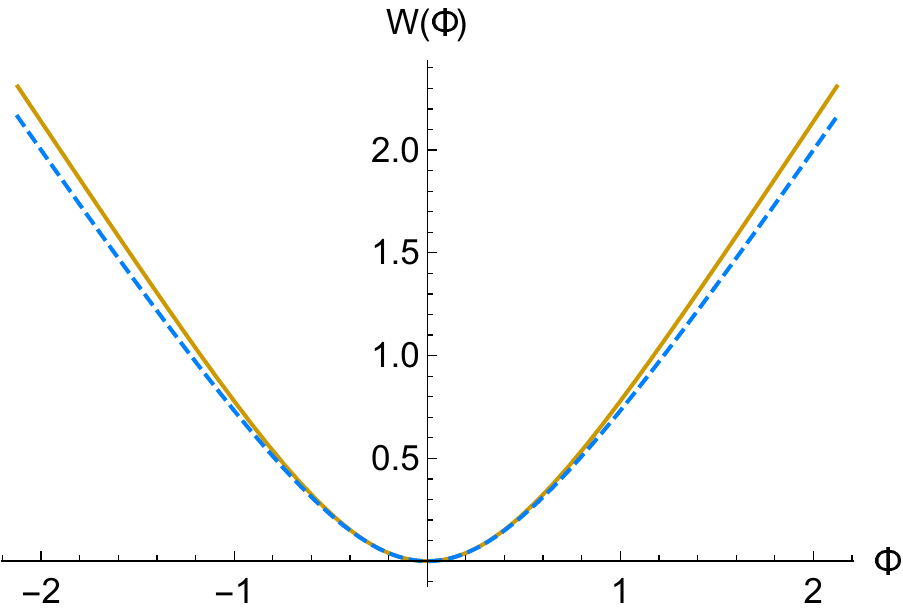}}~
\subcaptionbox{The potentials of the {\it log cosh}-type.\label{sfig:LogCosh}}{\includegraphics[width=0.47\textwidth]{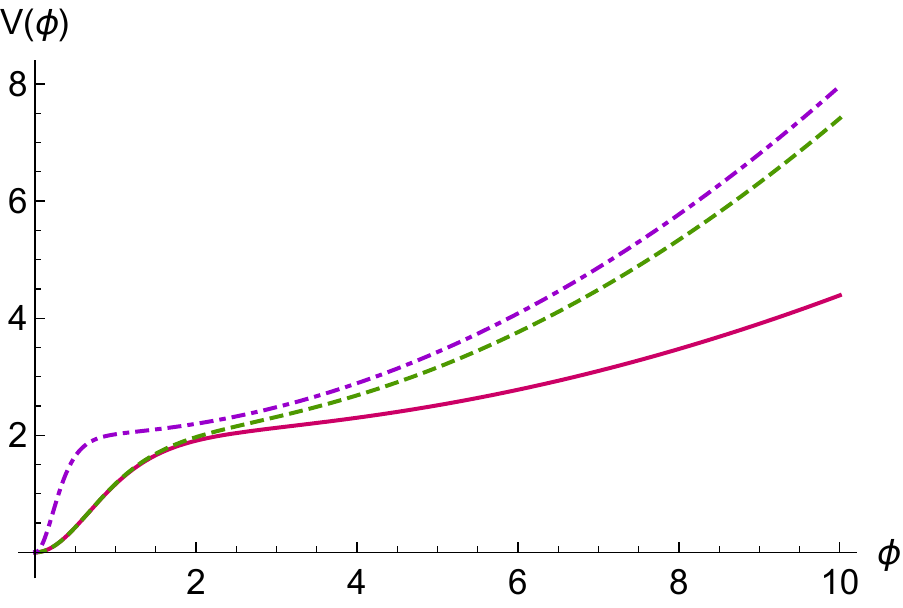}}
  \caption{
 The examples of asymptotically linear superpotentials and the corresponding potentials.
(\ref{sfig:aLinearW})  The yellow solid (blue dashed) line corresponds to $W=\log ( \cosh \sqrt{2}\Phi )$ ($W=\sqrt{1+2\Phi^2}-1$).
  (\ref{sfig:LogCosh}) The potential~\eqref{Vquad}.  The red solid, green dashed, and purple dot-dashed lines correspond to $(c, a)=(1.74, 1), (1.75, 1),$ and $(1.75, 3)$, respectively, with $m=1/a$. 
  }
  \end{figure}

If the value of $c$ is close to the critical value $\sqrt{3}$, the potential is dominated by the derivative term.
To obtain an asymptotically flat potential, the superpotential has to approach a linear function asymptotically.
Besides, the value and the slope of the superpotential at the origin should vanish.
Those superpotentials are shown in Fig.~\ref{sfig:aLinearW}.

A good example is given by
\paragraph{Type 2a: the {\it log cosh} model}
\begin{align}
W=m \left( \log \left(  \cosh  \sqrt{2} a \Phi  \right) \right),
\end{align}
where $a$ is a real parameter, $m$ sets the scale of the inflationary potential.  The leading-order scalar potential~\eqref{Vleading} takes the form
\begin{align}
V/m^2 = a^2 \tanh^2 a\phi+ \left( c^2 -3 \right) \left(\log \left( \cosh a \phi \right)\right)^2. \label{Vquad}
\end{align}
This potential is shown in Fig.~\ref{sfig:LogCosh}.
When the second term is negligible, the potential is that of the T-model~\cite{Kallosh:2013hoa, Kallosh:2015lwa}.
When the second term dominates, the potential becomes a quadratic function. It is also possible to interpolate between 
the plateau potential and the quadratic one, as the limiting cases. Taking the small $a$ limit, we get the potential that  is
close to a quartic one near the origin. Those potentials have a rich structure depending on the values of the parameters.

Yet another case is
\paragraph{Type 2b: the square-root-model}  \mbox{}\\
with
\begin{align}
W=m \left( \sqrt{1+ 2 a^2 \Phi^2} -1 \right),
\end{align}
which are similar to the previous type-2a (log cosh) model, see Fig.~\ref{sfig:aLinearW}.
The potential~\eqref{Vleading} in this case reads
\begin{align}
V/m^2 =\frac{2 a^4 \phi^2}{1+a^2 \phi^2} + (c^2 -3) \left( a^2 \phi^2 -2 \sqrt{1+a^2 \phi^2} + 2 \right).
\end{align}

The last two types have a singularity and/or a branch cut off the inflationary trajectory. There exist infinitely many superpotentials with similar predictions in the field region relevant to observations.

Next, we consider the case when $c$ is sufficiently large so that the non-derivative term dominates and asymptotes to a constant in the large-field region. For example, it is well represented by the following model:
\paragraph{Type 3a: the {\it tanh} model} \mbox{}\\
 with
\begin{align}
W=m \tanh^2 \sqrt{2} a\Phi~,
\end{align}
where $m$ sets the scale of inflation, and $a$ is a real parameter.
The potential~\eqref{Vleading} becomes
\begin{align}
V/m^2 = \tanh^2 a\phi \left[ \frac{4a^2}{\cosh^2 a\phi} + \left(c^2 -3 \right) \tanh^2 a\phi \right]~.\label{Vtanh}
\end{align}
It yields a flat potential for inflation, $V\sim \tanh^4 a\phi$, but in the $c\to\sqrt{3}$ limit it does not lead to a plateau potential.  The small $a$ limit leads to a quartic potential. This is because the first term in the expansion of the SUSY-restoring superpotential~\eqref{Wexpansion} is quadratic, while the main part of the scalar potential is proportional to its square.

One may also expand an asymptotically constant superpotential as a constant plus a series of decaying functions.

\paragraph{Type 3b: the models with exponentials}\hspace{-4mm},
\begin{align}
W=a_0 + a_1 e^{-b_1 \Phi} + a_2 e^{-b2 \Phi}.
\end{align}
This case was studied in Section~\ref{sec:analyses} in detail.

Finally, we consider the following type of models.
\paragraph{Type 3c: the models with a rational function}\hspace{-4mm},
\begin{align}
W = \frac{a_2 \Phi^2 + a_3 \Phi^3 + \cdots + a_n \Phi^n}{1 + b_1 \Phi + b_2 \Phi^2 + \cdots + b_m \Phi^m}.
\end{align}
This can be viewed as the Pad\'{e} approximation of order [$n$/$m$] of some holomorphic function.
The numerator begins with the quadratic term to satisfy~\eqref{SUSYP}.
Here, we take $n=m$ to obtain an asymptotically constant potential, and set $n=2$ as the simplest choice.
Then, the leading-order potential is
\begin{align}
V / a_2^2 = \frac{4(2\sqrt{2}+b_1 \phi )^2 \phi^2}{(2+ \sqrt{2} b_1 \phi + b_2 \phi^2)^4} + (c^2 -3) \frac{\phi^4}{(2+\sqrt{2}b_1 + b_2 \phi^2 )^2 } .
\end{align}
To further simplify the model, consider the case of $b_1 = 0$.
The asymptotic form of the potential is a constant plus a fall-off like $\phi^{-2}$.
This inverse-hilltop potential yields $n_{\text{s}} \simeq 1 - \frac{3}{2N}$ and $r \simeq 2 \sqrt{\frac{2}{b_2}}\frac{1}{N^{3/2}}$.

Needless to say, our classification here is incomplete, being the first step in that direction.

\bibliographystyle{utphys}
\bibliography{SSI_SUSY2.bib}
\end{document}